\documentclass[10pt,conference]{IEEEtran}
\IEEEoverridecommandlockouts

\usepackage{cite}
\usepackage{amsmath,amssymb,amsfonts}
\usepackage{algorithmic}
\usepackage[linesnumbered,ruled]{algorithm2e}
\usepackage{graphicx}
\usepackage{textcomp}
\usepackage{xcolor}
\usepackage{comment}
\usepackage{booktabs}
\usepackage{listings}
\usepackage{mdframed}
\usepackage{cleveref}
\usepackage{breqn}
\usepackage{stfloats}
\usepackage{afterpage}
\usepackage{float}
\usepackage{caption}
\usepackage{subcaption}
\usepackage{placeins}
\usepackage{enumitem}
\usepackage{url}

\usepackage{tikz}
\usepackage{lipsum} 
\makeatletter
\def\ieeecopyright{
  \footnotesize
  © 2025 IEEE. Personal use of this material is permitted.\newline
  DOI: 10.1109/RTAS65571.2025.00017}
\makeatother
\AddToHook{shipout/firstpage}{%
  \begin{tikzpicture}[remember picture,overlay]
    \node[anchor=south west,xshift=1.0cm,yshift=0.8cm]
      at (current page.south west)%
      {\parbox{\linewidth}{\raggedright\ieeecopyright}};
  \end{tikzpicture}%
}

\def\BibTeX{{\rm B\kern-.05em{\sc i\kern-.025em b}\kern-.08em
    T\kern-.1667em\lower.7ex\hbox{E}\kern-.125emX}}

\colorlet{codebackground}{gray!3}
\colorlet{codeborder}{black}

\crefname{figure}{Fig.}{Figures}
\crefname{table}{TABLE}{Tables}
\crefname{section}{Section}{Sections}
\crefname{equation}{eq.}{eqs.}

\lstdefinelanguage{yaml}{
  basicstyle=\small\ttfamily,
  morestring=[b]',
  morestring=[b]",
  morecomment=[l]\#,
  morekeywords={true,false,null,y,n},
  keywordstyle=\color{blue},
  stringstyle=\color{red},
  commentstyle=\color{green},
  backgroundcolor=\color{codebackground},
  frame=single, 
  rulecolor=\color{codeborder},
  framesep=3pt, 
  rulesepcolor=\color{codeborder},
  xleftmargin=4pt,
  xrightmargin=4pt,
  rulesep=5pt,
  breaklines=true,  
  breakindent=10pt, 
  postbreak=\mbox{\textcolor{gray}{$\hookrightarrow$}\space} 
}

\newcommand{\TOOLNAME}{\textit{CallbackIsolatedExecutor}}

\begin{document}

\renewcommand{\thefootnote}{\fnsymbol{footnote}}

\title{
Work in Progress:
Middleware-Transparent Callback Enforcement in Commoditized \\Component-Oriented Real-time Systems
}

\author{
  Takahiro Ishikawa-Aso$^{\dagger\ddagger}$,
  Atsushi Yano$^{*\ddagger}$,
  Takuya Azumi$^{*}$,
  Shinpei Kato$^{\dagger}$
  \\
  $^{\dagger}$The University of Tokyo, Japan \space
  $^{*}$Saitama University, Japan \\
  $^{\ddagger}$TIER IV Incorporated, Japan
}

\maketitle

\begin{abstract}
Real-time scheduling in commoditized component-oriented real-time systems, such as ROS~2 systems on Linux, has been studied under nested scheduling: OS thread scheduling and middleware layer scheduling (e.g., ROS~2 Executor). 
However, by establishing a persistent one-to-one correspondence between callbacks and OS threads, we can ignore the middleware layer and directly apply OS scheduling parameters (e.g., scheduling policy, priority, and affinity) to individual callbacks.
We propose a middleware model that enables this idea and implements \TOOLNAME{} as a novel ROS~2 Executor.
We demonstrate that the costs (user-kernel switches, context switches, and memory usage) of \TOOLNAME{} remain lower than those of the \textit{MultiThreadedExecutor}, regardless of the number of callbacks.
Additionally, the cost of \TOOLNAME{} relative to \textit{SingleThreadedExecutor} stays within a fixed ratio (1.4x for inter-process and 5x for intra-process communication).
Future ROS~2 real-time scheduling research can avoid nested scheduling, ignoring the existence of the middleware layer.
\end{abstract}

\begin{IEEEkeywords}
Middleware,
Publish-subscribe,
Robot programming,
Real-time systems,
Scheduling algorithms,
Multithreading
\end{IEEEkeywords}

\vspace{-6mm}

\section{Introduction}
\begin{figure}[tb]
  \centering
  \includegraphics[width=\linewidth]{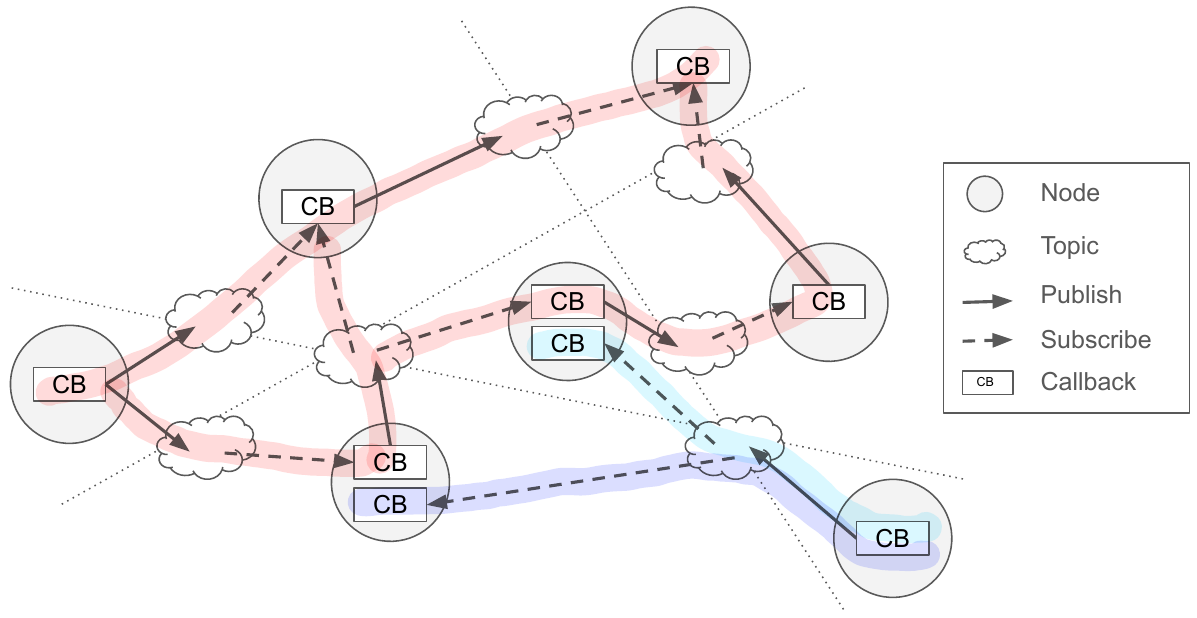}
  \vspace{-7mm}
  \caption{Component-oriented real-time system (e.g., ROS~2).} 
  \label{fig-component-oriented-realtime}
  \vspace{-6mm}
\end{figure}

\vspace{-1mm}
Many modern autonomous cyber‐physical systems (CPS), such as autonomous driving and robotics systems, are modeled as component‐oriented real‐time systems (\cref{fig-component-oriented-realtime}).
A typical system comprises independent nodes that mutually publish and subscribe to messages, collectively creating integrated dataflows throughout the system.
Nodes operate independently by either publishing to or subscribing to topics defined by specific data types and identifiers.
Due to the extensive ecosystem, they are typically deployed on the most commoditized platform: Robot Operating System~2 (ROS~2) \cite{ros2_repo} on Linux.

Due to the real-time interaction with the physical world, autonomous CPS requires temporal correctness as well as functional correctness.
The temporal constraints on the dataflows are defined in terms of \textit{freshness}, \textit{consistency}, and \textit{stability} \cite{li2024data}.
Research on real-time scheduling in ROS~2 has primarily targeted the optimization of end-to-end \textit{response time}, a metric similar to freshness.
In this context, cause-effect chains with ROS~2 semantics incorporated have been commonly employed as a model~\cite{teper2022end, teper2023timing, teperend}.
There are also studies where systems are modeled as Directed Acyclic Graphs (DAGs) \cite{blass2021automatic, partap2022device}.

\begin{figure*}[tb]
  \centering
  \begin{minipage}[b]{0.35\linewidth}
    \includegraphics[width=\linewidth]{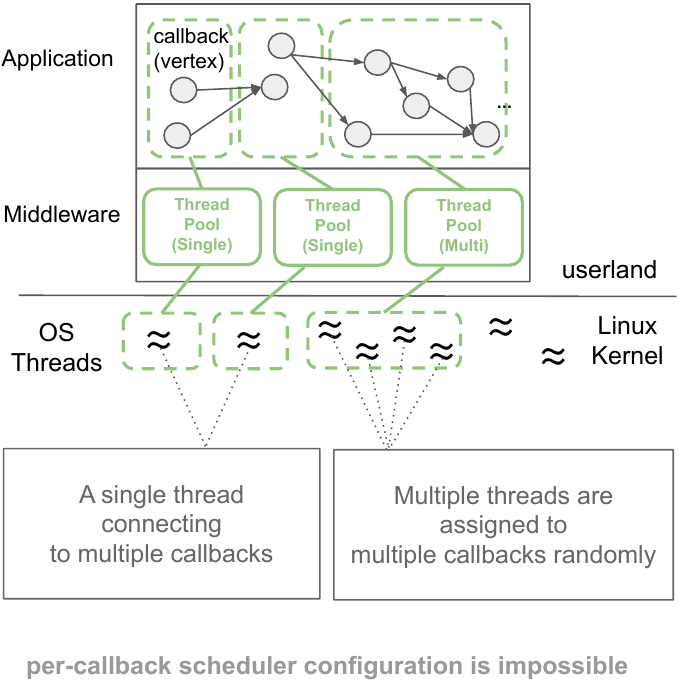}
    \vspace{1.8mm}
    \caption*{(a) Existing Executors' callback enforcement}
  \end{minipage}
  \hfill
  \begin{minipage}[b]{0.64\linewidth}
    \includegraphics[width=\linewidth]{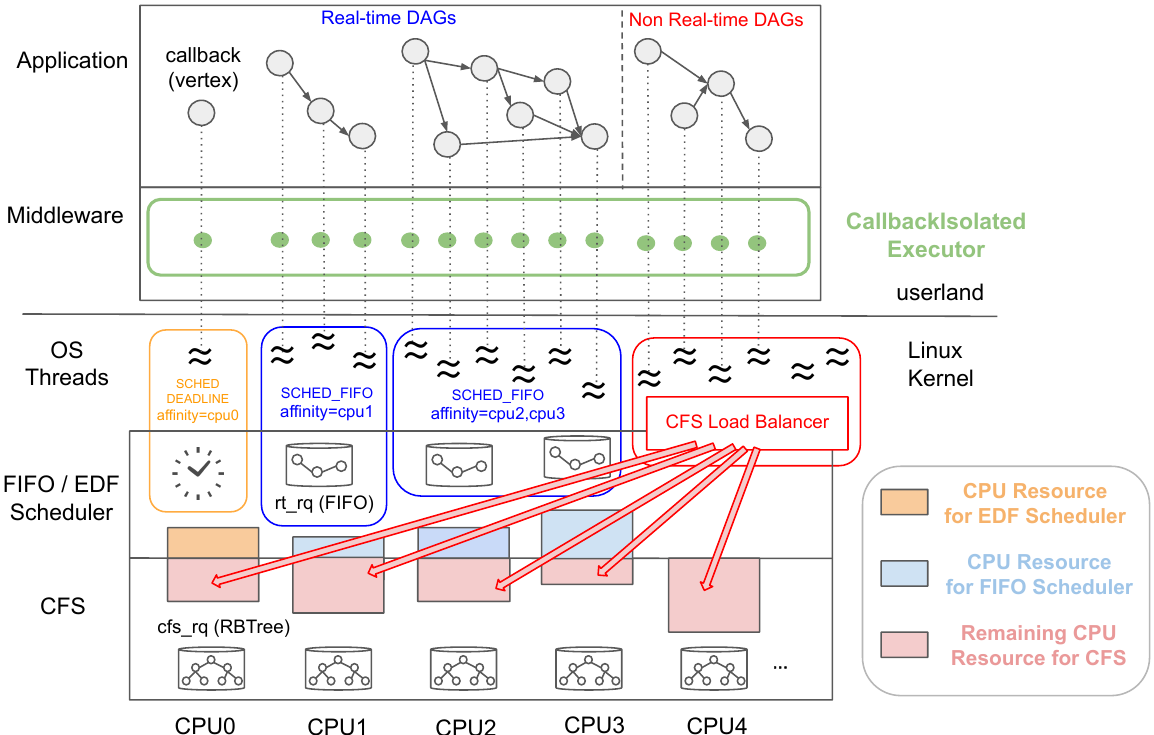}
    \caption*{(b) Proposed \textit{CallbackIsolatedExecutor's} callback enforcement}
  \end{minipage}
    \caption{Middleware design of existing Executors and \TOOLNAME{} in ROS~2 applications on Linux.}
    \label{fig-system-model}
    \vspace{-3mm}
\end{figure*}

In both models, the real-time community has assumed nested scheduling (\cref{fig-system-model}~(a)) where both the callback scheduler and the thread scheduler are considered.
This approach has led to unnecessarily complex response time analysis and scheduling algorithms \cite{casini2019response, tang2020response, choi2021picas, blass2021ros, choi2022priority, teper2022end, jiang2022real, chaaban2023new, tang2023real, sobhani2023timing, al2024dynamic, teperend, teper2024thread}.
The nested scheduling is an artificial complexity imposed by the ROS~2 implementation rather than an inherent necessity.
This complexity has prevented the effective utilization of traditional scheduling techniques.
An effort of extending the \textit{EventsExecutor} addressed this issue~\cite{teper2024bridging}, but it provides only limited functionality.

We propose a middleware design that enables direct configuration of OS scheduling parameters (scheduler, scheduler-specific parameters, and affinity) for each callback, bypassing nested scheduling.
Consequently, real-time scheduling in ROS~2 no longer needs to account for the presence of the Executor in theoretical models.
Instead, focus can be placed exclusively on the callback dependency graph and enforcement of the scheduling parameters on callback vertices.
We develop \textbf{\TOOLNAME{}} and the corresponding \textit{Component Container} with the one-to-one correspondence between callbacks and OS threads.
\TOOLNAME{} eliminates the need for future research to address the nested scheduling problem in ROS~2 applications.
Furthermore, we evaluate the overhead of \TOOLNAME{} compared to \textit{SingleThreadedExecutor} and \textit{MultiThreadedExecutor}.

\textbf{Contributions.} The main contributions of this paper include the following:
\vspace{-1mm}
\begin{enumerate}
\item We propose a middleware design that eliminates nested scheduling by establishing a persistent one-to-one correspondence between callbacks and OS threads.
\item We formulate the scheduling problem of ROS~2 on Linux so the Executor layer can be ignored.
\item We demonstrate that \TOOLNAME{} keeps acceptable overheads compared to existing Executors.
\end{enumerate}

\section{New ROS~2 Executor Design} \label{section-system-model}
\vspace{-3mm}
\textbf{Essence of ROS~2 Real-time Scheduling}.
A ROS~2 application comprises DAGs formed by callback chains. 
Dependencies between DAGs via shared variables in nodes are incorporated into the cause-effect model. 
A recent study \cite{rtss2024_yano_wip} proposes setting independent deadlines for each DAG, as applying cause-effect chains to complex ROS~2 applications, such as Autoware \cite{kato2015open, kato2018autoware}, is impractical.
Regardless of the modeling approach, the scheduling problem of ROS~2 systems fundamentally concerns how CPU resources are allocated to callbacks, which act as subtasks in DAG scheduling.

\textbf{Artificial Complexity in Scheduling Imposed by ROS~2's Implementation}.
In ROS~2 systems, scheduling parameters should ideally be applied directly to each individual callback (subtask).
However, due to the implementation constraints of the existing ROS~2 Executors, this is not feasible (\cref{fig-system-model}(a)).
ROS~2 provides two primary types of Executors: the \textit{SingleThreadedExecutor} and the \textit{MultiThreadedExecutor}.
In the \textit{SingleThreadedExecutor}, multiple callbacks are assigned to a single OS thread.
Since scheduling parameters are set at the OS thread level, they cannot be configured per callback in this model.
In contrast, the \textit{MultiThreadedExecutor} employs multiple OS threads and dispatches callbacks in a many-to-many relationship, with each invocation randomly assigned to a thread.
As a result, specific scheduling parameters cannot be assigned to individual callbacks.
Both Executors adopt a ROS~2-specific round-robin-like scheduling policy, hindering the direct application of classical real-time and DAG scheduling theories.
Consequently, ROS~2 Executor-specific scheduling theories have been studied in recent years.
Unlike existing Executors that poll a \textit{wait-set}, the \textit{EventsExecutor} pushes executable callbacks directly into a FIFO queue upon event occurrence.
However, it also encounters nested scheduling.

\textbf{Ignore the Executor, Focus on the Callback Graph}.
To configure scheduling parameters on a per-callback basis, we propose a novel middleware design where each callback has a dedicated OS thread (\cref{fig-system-model}~(b)).
Since each OS thread corresponds to a single callback, scheduling parameters can be applied by setting them on the corresponding thread.
If a callback never gets ready for the next period until the previous execution finishes (i.e., its minimum inter-arrival time exceeds its worst-case turnaround time), the middleware layer can be ignored in scheduling.
Thus, scheduling reduces to applying OS parameters to DAG vertices, where vertices are callbacks and edges denote execution dependencies.
The only concern with \TOOLNAME{} is the overhead introduced by maintaining a one-to-one mapping between callbacks and OS threads.
This issue is discussed and evaluated in \cref{section-evaluation}.

\vspace{-1mm}
\section{Proposed ROS~2 Scheduling Problem} \label{section-scheduling-problem}
\vspace{-1mm}
With \TOOLNAME{}, the scheduling in ROS~2 applications can be modeled without considering the presence of the middleware layer.
The proposed scheduling problem formulation on ROS~2 Linux assumes the following two constraints.
We should design ROS~2 applications accordingly.

\begin{figure}[tb]
  \centering
  \includegraphics[width=\linewidth]{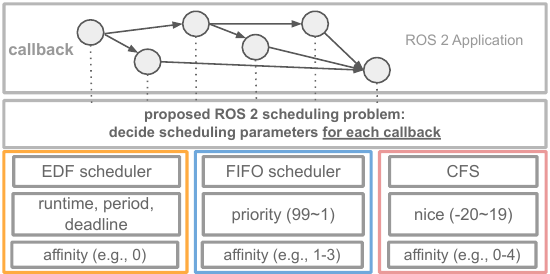}
  \caption{Proposed formulation of ROS~2 scheduling enforcement. We directly specify scheduler, scheduler-specific parameters, and core affinity for each callback.}
  \vspace{1mm}
  \label{fig-proposed-scheduling-problem}
  \vspace{-3mm}
\end{figure}

\begin{itemize}[leftmargin=15pt]
\item \textbf{Each \textit{CallbackGroup} Contains One Callback}.
In ROS~2, \textit{CallbackGroup} is a mechanism for grouping callbacks intended not to be executed in parallel.
The previous section discussed a one-to-one mapping between callbacks and OS threads, but in practice, \textit{CallbackGroups}, not callbacks, are mapped to threads.
To implement \cref{fig-system-model}~(b), each callback must belong to a separate \textit{CallbackGroup}.
The primary purpose of a \textit{CallbackGroup} is to facilitate mutual exclusion for shared resources, but they should fundamentally be managed using mutexes. 

\item \textbf{Callbacks are Not Reentrant}.
As mentioned in \cref{section-system-model}, the assumption ``a callback never gets ready for the next period until the previous execution finishes'' must hold.
Many practical ROS~2 applications do not implement reentrant callbacks, making this constraint reasonable.
If this assumption does not hold (e.g., when a callback has a long execution time), the callback can be split into multiple smaller callbacks to enable pipeline processing.
\end{itemize}
Under these two constraints, scheduling enforcement in ROS~2 on Linux follows \cref{fig-proposed-scheduling-problem}, where each callback is assigned a scheduling policy, policy-specific parameters, and core affinity (i.e., which CPU core it runs on).
This formulation aligns with classical real-time and DAG scheduling theories, enabling the leverage of existing research assets.

Linux provides three types of schedulers, ranked in order of priority: Earliest Deadline First (EDF), FIFO, and Completely Fair Scheduler (CFS).
Each scheduler operates independently, such that higher-priority schedulers may preempt or affect lower-priority ones, but not vice versa.
Leveraging this property, as illustrated in \cref{fig-system-model}~(b), the unused CPU time from EDF and FIFO schedulers can be fully exploited by CFS.
Real-time DAGs run on an EDF scheduler or FIFO scheduler, while non-real-time DAGs and other threads use CFS, ensuring real-time scheduling and maximizing CPU utilization.
Large-scale ROS~2 applications, such as Autoware, often involve hundreds of nodes. Additionally, efficient CPU utilization is crucial for power efficiency, making this approach essential.
The characteristics of each scheduler are described below:
\begin{itemize}[leftmargin=15pt]
\item \textbf{EDF scheduler:} In Linux, the EDF scheduler is combined with the Constant Bandwidth Server (CBS). Originally designed for traditional periodic tasks, it may have potential applications for DAG scheduling.
\item \textbf{FIFO scheduler:} A preemptive static priority scheduler. Scheduling of callback graphs on this scheduler is formulated as fixed-priority preemptive DAG scheduling. It is a primary option for DAG scheduling.
\item \textbf{CFS:} A scheduler that aims to achieve fair CPU time distribution by prioritizing tasks with the least virtual runtime (vruntime). The nice value reduces vruntime growth by about 10\% per decrement, increasing a task’s CPU share.
\end{itemize}

\begin{figure}[h]
  \centering
  \includegraphics[width=\linewidth]{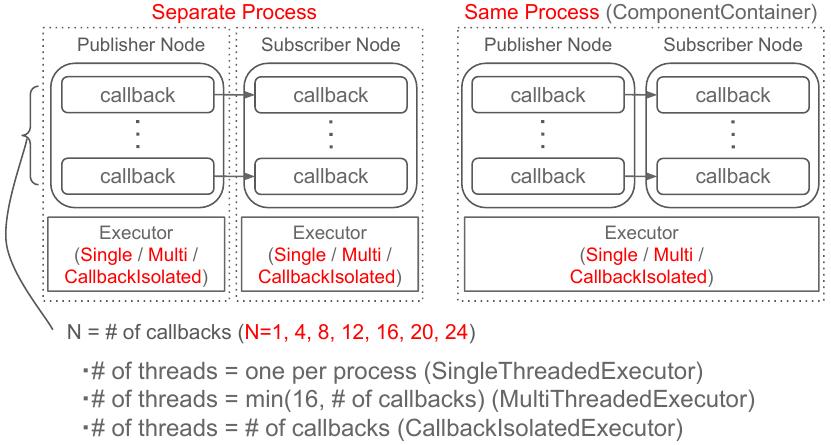}
  \caption{Experimental conditions.}
  \label{fig-evaluation-diagram}
  \vspace{-5mm}
\end{figure}

\section{Overhead of Callback Isolated Executor} \label{section-evaluation}
From the perspective of theoretical real-time scheduling models, \TOOLNAME{} is a superior choice, offering only advantages.
However, the overhead of mapping each callback to a dedicated thread must be discussed.
We measure user-kernel switches, context switches, and memory consumption.
Additionally, we discuss \textit{wait-set} contention due to thread count increases.
The evaluation platform uses an Intel Xeon E-2278GE (3.30 GHz, 8 cores, 16 threads with Hyper-Threading), 32 GB RAM, Ubuntu 22.04 LTS, Linux kernel 6.8, and ROS 2 Humble with Cyclone DDS.

To evaluate overheads of \TOOLNAME{} against existing Executors, we prepare a ROS~2 node pair with publish-subscribe communication with the varied experimental conditions shown in \cref{fig-evaluation-diagram}.
Each publisher and subscriber node holds N callbacks (N = 1, 4, .., 24), each in a separate CallbackGroup, maintaining a one-to-one mapping.
The publisher node runs a timer callback that publishes every 10 ms, while the subscriber node executes a subscription callback that receives messages.
We compare the impact of N on overhead across \textit{SingleThreadedExecutor}, \textit{MultiThreadedExecutor}, and \TOOLNAME{}.
To examine inter-process vs. intra-process differences, we run publisher and subscriber nodes either in separate or shared processes.
For inter-process communication conditions, we plot the sum of the measurements from the publisher and subscriber node processes.

\begin{figure*}[tb]
  \centering
  \begin{minipage}[b]{0.329\linewidth}
    \includegraphics[width=\linewidth]{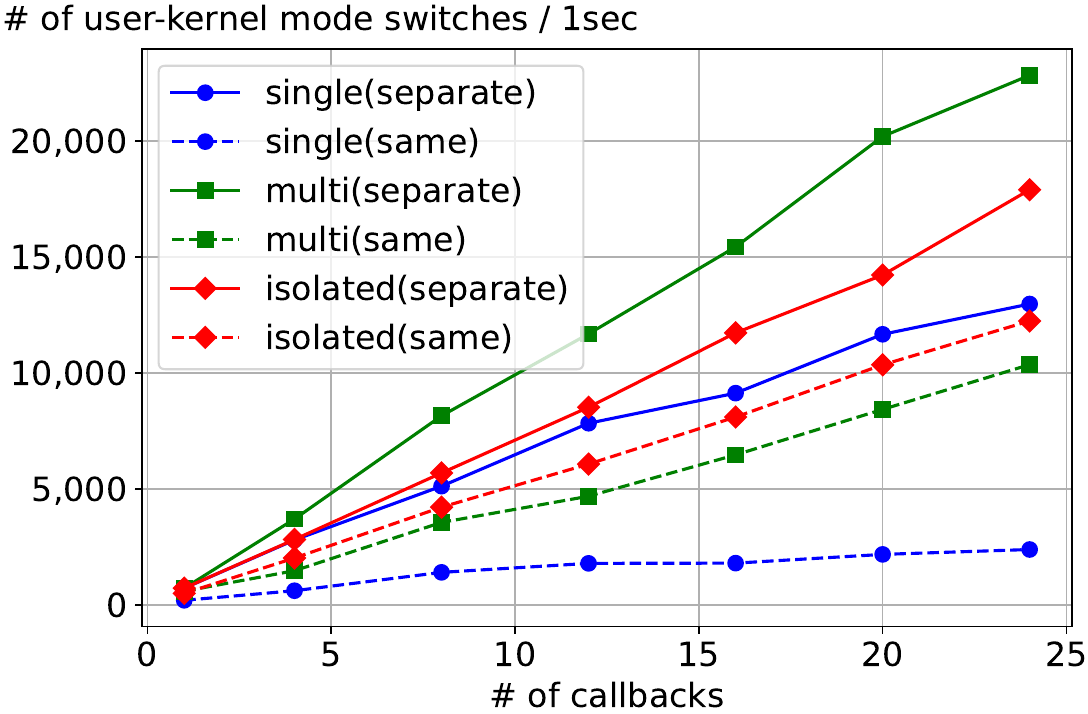}
    \caption*{(a) \# of user-kernel switches per sec.}
  \end{minipage}
  \hfill
  \begin{minipage}[b]{0.329\linewidth}
    \includegraphics[width=\linewidth]{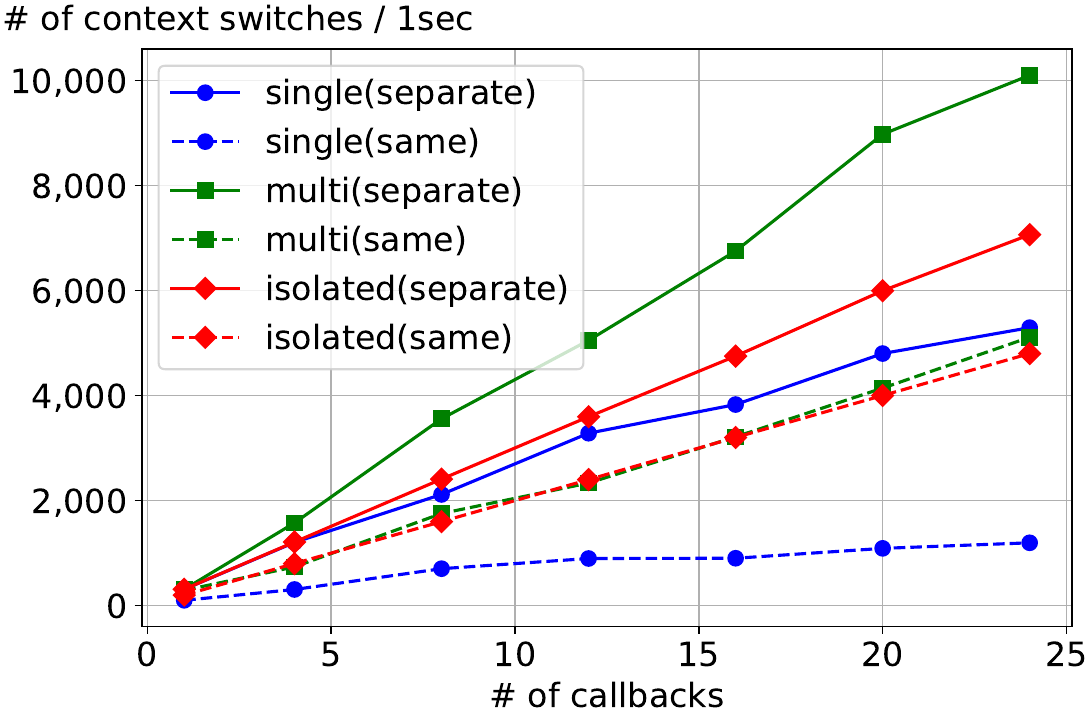}
    \caption*{(b) \# of context switches per sec.}
  \end{minipage}
  \hfill
  \begin{minipage}[b]{0.329\linewidth}
    \includegraphics[width=\linewidth]{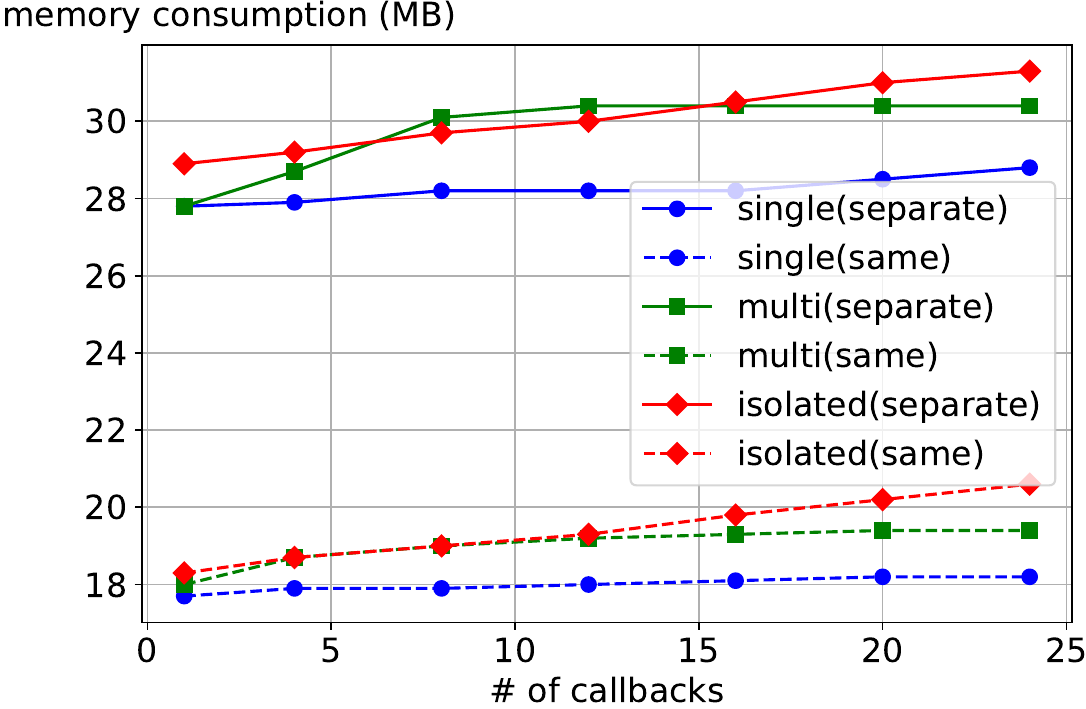}
    \caption*{(c) Memory consumption}
  \end{minipage}
    \caption{Comparison of \TOOLNAME{} overhead with existing Executors (see \cref{fig-evaluation-diagram} for experimental conditions)}
    \label{fig-evaluation}
    \vspace{-4mm}
\end{figure*}

\cref{fig-evaluation} shows the impact of callback count on overhead factors for each Executor.
Memory consumption in \TOOLNAME{} is slightly higher but negligible.
Compared to \textit{MultiThreadedExecutor}, despite \textit{MultiThreadedExecutor} being limited by hardware concurrency, \TOOLNAME{} outperforms it in all but user-kernel mode switches in the same process.
The slight advantage of \textit{MultiThreadedExecutor} in user-kernel mode switches under intra-process execution is likely due to pointer passing communication, allowing to bypass kernel-space operations.
Compared to \textit{SingleThreadedExecutor}, \TOOLNAME{} maintains user-kernel mode switches and context switches within a fixed ratio.
These switches remain within about 5× for intra-process publish-subscribe and 1.4× for inter-process communication.
This ratio is bounded regardless of the number of callbacks.

In \textit{MultiThreadedExecutor}, multiple threads share one \textit{wait-set}, incurring overhead from reconstructing the \textit{wait-set} and acquiring locks.
In contrast, \TOOLNAME{} does not share \textit{wait-set} across threads, eliminating this overhead.

\vspace{-0mm}

\section{Conclusion and Future Work}
With \TOOLNAME{}, the real-time community no longer needs to account for the existence of the Executor in ROS~2 scheduling \textbf{in practical cases}.
This is because real-world ROS~2 systems rarely have dozens of callbacks per node, and the opportunistic locking mechanisms based on futexes in Linux tend to scale well with a large number of threads and low levels of contention.
As future work, we plan to evaluate our approach on real-world systems such as Autoware, and to explore concrete scheduling techniques tailored to such systems.
Scheduling research will advance based on our formulation proposed for ROS~2 applications on Linux.
\TOOLNAME{} is available as open-source\footnote{\url{ https://github.com/sykwer/callback_isolated_executor.}}.

\section{Acknowledgment}
This research is based on results obtained from Green Innovation Fund Projects / Development of In-vehicle Computing and Simulation Technology for Energy Saving in Electric Vehicles, JPNP21027, subsidized by the New Energy and Industrial Technology Development Organization (NEDO) and partially by JST PRESTO Grant Number JPMJPR21P1.

\vskip\baselineskip

\bibliographystyle{IEEEtran}
\bibliography{references}

\end{document}